\documentclass[aps,prl,preprint,tightenlines,superscriptaddress,showpacs,byrevtex]{revtex4}
\usepackage{graphicx} 
\usepackage{dcolumn}  

\graphicspath{{ps}}

\begin{document}



\title{ \quad\\[1.0cm] Measurement of the $\tau$ lepton mass \\
and an upper limit on the
mass difference between $\tau^+$ and $\tau^-$}

\affiliation{Budker Institute of Nuclear Physics, Novosibirsk}
\affiliation{Chonnam National University, Kwangju}
\affiliation{University of Cincinnati, Cincinnati, Ohio 45221}
\affiliation{Department of Physics, Fu Jen Catholic University, Taipei}
\affiliation{The Graduate University for Advanced Studies, Hayama, Japan}
\affiliation{University of Hawaii, Honolulu, Hawaii 96822}
\affiliation{High Energy Accelerator Research Organization (KEK), Tsukuba}
\affiliation{University of Illinois at Urbana-Champaign, Urbana, Illinois 61801}
\affiliation{Institute of High Energy Physics, Chinese Academy of Sciences, Beijing}
\affiliation{Institute of High Energy Physics, Vienna}
\affiliation{Institute of High Energy Physics, Protvino}
\affiliation{Institute for Theoretical and Experimental Physics, Moscow}
\affiliation{J. Stefan Institute, Ljubljana}
\affiliation{Kanagawa University, Yokohama}
\affiliation{Korea University, Seoul}
\affiliation{Kyungpook National University, Taegu}
\affiliation{Swiss Federal Institute of Technology of Lausanne, EPFL, Lausanne}
\affiliation{University of Ljubljana, Ljubljana}
\affiliation{University of Maribor, Maribor}
\affiliation{University of Melbourne, Victoria}
\affiliation{Nagoya University, Nagoya}
\affiliation{Nara Women's University, Nara}
\affiliation{National Central University, Chung-li}
\affiliation{National United University, Miao Li}
\affiliation{Department of Physics, National Taiwan University, Taipei}
\affiliation{H. Niewodniczanski Institute of Nuclear Physics, Krakow}
\affiliation{Nippon Dental University, Niigata}
\affiliation{Niigata University, Niigata}
\affiliation{University of Nova Gorica, Nova Gorica}
\affiliation{Osaka City University, Osaka}
\affiliation{Osaka University, Osaka}
\affiliation{Panjab University, Chandigarh}
\affiliation{Peking University, Beijing}
\affiliation{RIKEN BNL Research Center, Upton, New York 11973}
\affiliation{University of Science and Technology of China, Hefei}
\affiliation{Seoul National University, Seoul}
\affiliation{Shinshu University, Nagano}
\affiliation{Sungkyunkwan University, Suwon}
\affiliation{University of Sydney, Sydney NSW}
\affiliation{Tata Institute of Fundamental Research, Bombay}
\affiliation{Toho University, Funabashi}
\affiliation{Tohoku Gakuin University, Tagajo}
\affiliation{Tohoku University, Sendai}
\affiliation{Department of Physics, University of Tokyo, Tokyo}
\affiliation{Tokyo Institute of Technology, Tokyo}
\affiliation{Tokyo Metropolitan University, Tokyo}
\affiliation{Tokyo University of Agriculture and Technology, Tokyo}
\affiliation{Virginia Polytechnic Institute and State University, Blacksburg, Virginia 24061}
\affiliation{Yonsei University, Seoul}
  \author{K.~Belous}\affiliation{Institute of High Energy Physics, Protvino} 
  \author{M.~Shapkin}\affiliation{Institute of High Energy Physics, Protvino} 
  \author{A.~Sokolov}\affiliation{Institute of High Energy Physics, Protvino} 
  \author{K.~Abe}\affiliation{High Energy Accelerator Research Organization (KEK), Tsukuba} 
  \author{K.~Abe}\affiliation{Tohoku Gakuin University, Tagajo} 
  \author{I.~Adachi}\affiliation{High Energy Accelerator Research Organization (KEK), Tsukuba} 
  \author{H.~Aihara}\affiliation{Department of Physics, University of Tokyo, Tokyo} 
  \author{D.~Anipko}\affiliation{Budker Institute of Nuclear Physics, Novosibirsk} 
  \author{K.~Arinstein}\affiliation{Budker Institute of Nuclear Physics, Novosibirsk} 
  \author{V.~Aulchenko}\affiliation{Budker Institute of Nuclear Physics, Novosibirsk} 
% ...
\collaboration{The Belle Collaboration}

\begin{abstract}
The mass of the $\tau$ lepton has been measured in the decay mode
$\tau \rightarrow 3\pi \nu_\tau$
using a pseudomass technique. The result obtained from
$414~\mathrm{fb}^{-1}$
of data collected with the Belle detector is
$M_\tau = (1776.61\pm 0.13 \mbox{(stat.)} \pm 0.35 \mbox{(sys.)})$
MeV/$c^2$.
The upper limit on the relative mass difference between positive and
negative $\tau$ leptons is
$|M_{\tau^+}-M_{\tau^-}|/M_{\tau} < 2.8 \times 10^{-4}$
at 90\% confidence level.
\end{abstract}
\maketitle

\tighten

{\renewcommand{\thefootnote}{\fnsymbol{footnote}}}
\setcounter{footnote}{0}

Masses of quarks and leptons are fundamental parameters of the
Standard Model (SM).
High precision measurements
of the mass, lifetime and
the leptonic branching fractions of the $\tau$ lepton can be used to 
test the lepton universality hypothesis embedded in the SM. The present
PDG value of the
$\tau$ mass~\cite{PDG} is dominated  by the result of the BES 
Collaboration~\cite{BES} and has 
an accuracy of about 0.3 MeV/$c^2$. The same level of accuracy in $\tau$ mass
measurement was recently reported by the KEDR Collaboration~\cite{KEDR}. 
The data collected by the Belle experiment
allow a measurement 
with similar accuracy 
to the BES and KEDR experiments but with
different systematic uncertainties; the latter experiments analyse the
cross
section for $\tau$ pair production near threshold
while Belle measures the four-momenta of the visible $\tau$ decay products 
at a center-of-mass (c.m.) energy of $\sqrt{s}=10.58$ GeV.
Eventually, by combining these high precision measurements, we will
significantly improve the accuracy of the $\tau$ mass determination.

Separate measurements of the masses of the $\tau^+$ and $\tau^-$
leptons in Belle, allow us to test the CPT theorem, which
demands their equality.
A similar test was previously performed by OPAL at LEP~\cite{OPAL} 
with the result
$(M_{\tau^+}-M_{\tau^-})/M_{\tau}<3.0\times 10^{-3}$ at 90\% CL.

To measure the $\tau$ mass, we use a pseudomass technique that
was first employed by the ARGUS collaboration~\cite{ARGUS}. This technique 
relies on the reconstruction of the invariant mass and energy of the 
hadronic system in hadronic $\tau$ decays. The analysed variable is
\begin{equation}
 M_{\mathrm{min}}=\sqrt{M_X^2+2(E_{\mathrm{beam}}-E_X)(E_X-P_X)},
\end{equation}
which is less than or equal to the $\tau$ lepton mass. Here
$M_X$, $E_X$ and $P_X$ are the invariant mass, 
energy and absolute
value of the momentum, respectively, of the hadronic system
in $e^+e^-$  c.m. frame, and $E_{\mathrm{beam}}$ 
is the energy of the electron (or positron) in this  
frame. 
In the absence
of initial and final state radiation and assuming a perfect measurement of
the four-momentum of the hadronic system, the distribution of 
$M_{\mathrm{min}}$ extends up to 
and has a sharp edge at $M_\tau$.
Initial (ISR) and final (FSR) state radiation as well as the finite momentum
resolution of the detector smear this edge.  
We can use the edge position from a fit to the experimental
$M_{\mathrm{min}}$ distribution as an estimator of the $\tau$ mass,
since the background processes in the selected $\tau^+\tau^-$ sample
have a featureless $M_{\mathrm{min}}$ distribution near $M_\tau$.

The analysis presented here is based on 414 fb$^{-1}$ 
of data taken at the $\Upsilon$(4S) resonance ($\sqrt{s} = 10.58$ GeV) with 
the Belle
detector at the KEKB asymmetric-energy $e^+e^-$ collider~\cite{KEKB}.
A detailed description of the Belle detector is given
elsewhere~\cite{BELLE}. We mention here only the detector components essential
for the present analysis.

Charged tracks are reconstructed from hit information in a central drift
chamber (CDC) located in a 1.5~T solenoidal magnetic field. The $z$ axis of
the detector and the solenoid are aligned antiparallel to the positron beam. 
Track trajectory coordinates of the charged particles near the collision
point are provided by a silicon vertex detector (SVD). Photon detection and
energy measurement are performed with a CsI(Tl) electromagnetic calorimeter
(ECL). Identification of charged particles is based on the information from
the time-of-flight counters (TOF) and silica aerogel Cherenkov counters (ACC).
The ACC provides good separation between kaons and pions or muons at momenta
above 1.2 GeV. The TOF system consists of a barrel of 128 plastic
scintillation counters, and is effective in $K/\pi$ separation mainly for
tracks with momentum below 1.2 GeV. The lower energy tracks are also
identified using specific ionization $(dE/dx)$ measurements in the CDC.
Electrons are identified by combining information from the ECL, ACC, TOF and
CDC~\cite{LE}.
The magnet return yoke is instrumented to form the $K_L$ and muon detector
(KLM), which detects muon tracks~\cite{LMU} and provides trigger signals.
The responses from these detectors determine the likelihood $L_i$ of particle
type $i \in \{e,\, \mu,\, \pi,\, K,\, p\}$.
A charged particle is identified as an electron if the 
corresponding likelihood ratio~\cite{LE}, $P_e>0.9$ or if the electron 
mass hypothesis
has the highest probability. The electron efficiency for $P_e>0.9$ is
approximately 90\% for a single electron embedded into a hadronic event.
Charged particles are identified as muons if the corresponding muon 
likelihood  ratio~\cite{LMU} $P_\mu>0.8$. 
The muon detection efficiency for this measurement
is approximately 91\%. The corresponding 
likelihood ratio
cut for kaons and protons is 0.8. 
The kaon and proton identification efficiencies are about 80\%.
All charged tracks that are not
identified as an electron, muon, kaon or proton are treated as pions. 
The $K_S^0$ candidates 
are formed from pairs of charged tracks intersecting in a
secondary vertex more than 0.3 cm from the beam spot in the plane
transverse to the beam axis;
the $\chi^2$ of the vertex fit is required to be less than 11 and
the $\pi^+\pi^-$ invariant mass of the candidate is required to be
0.48 GeV/$c^2 < M_{\pi^+\pi^-} < 0.52$ GeV/$c^2$.
The $\pi^0$ candidates are formed from pairs of photons, each with 
energy greater than 0.1 GeV, that satisfy the condition  0.115 GeV/$c^2 < 
M_{\gamma \gamma} < 0.152 \mathrm{~GeV/}c^2$. 

We select events that have one $\tau$ lepton decaying leptonically
into $l\bar{\nu_l}\nu_\tau$ and the other into three charged 
pions and neutrino.
For the entire event, we require three charged pions and one
lepton (either muon or electron) with net charge equal to zero.
The number of charged kaons, protons, $K_S^0$ mesons and $\pi^0$
mesons should be equal to zero."
%
%
After applying all selection
criteria $5.8 \times 10^6$ events remain.

\begin{figure}
\vspace*{-8mm}
  \includegraphics[width=0.4\textwidth] {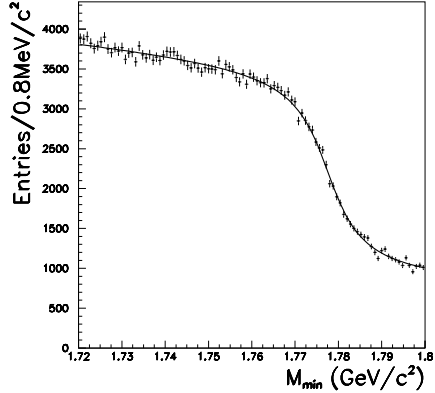}
\caption{The pseudomass distribution $M_{\mathrm{min}}$ for
the $\tau^\pm \rightarrow 3\pi^\pm \nu_\tau$ candidates. The points with error
bars are data and the solid line is the result of the fit with the
function (2). }
\label{fig1}
\vspace*{-3mm}
\end{figure}

The $M_{\mathrm{min}}$ distribution for the 
$\tau \rightarrow 3\pi \nu_\tau$ data
is shown in Fig. 1.
A fit was performed to these data with the empirical edge function
\begin{equation}
F(x) = (P_3+P_4 x) \mbox{arctan}((x-P_1)/P_2)+P_5+P_6 x,
\end{equation}
where $P_i$ are parameters of the fit. The fit range 1.72--1.80 GeV/$c^2$
is chosen.
The value of the uncorrected $\tau$ mass estimator, $P_1$, obtained 
from the fit is $P_1 = 1777.77 \pm 0.13$ MeV/$c^2$. 

To obtain the value of the $\tau$ mass from the $\tau$ mass estimator $P_1$
we use several Monte Carlo samples of $\tau^+\tau^-$ events where one $\tau$
decays leptonically and the other one decays into three charged pions and
neutrino. The KORALB generator~\cite{KORALB} is used for the Monte Carlo
$e^+e^- \rightarrow \tau^+\tau^-$ event production.

We use three different Monte Carlo samples with input $\tau$ masses equal to
1777.0~MeV/$c^2$, 1776.0~MeV/$c^2$ and 1776.8~MeV/$c^2$ for the first, second
and third sample, respectively. The statistics of each sample is approximately
equal to that of the data. The differences between the fitted estimator $P_1$
and the input $\tau$ mass for these samples are
$\Delta_1=(1.27\pm0.12)$~MeV/$c^2$, $\Delta_2=(1.29\pm0.05)$~MeV/$c^2$ and
$\Delta_3=(1.06\pm0.04)$~MeV/$c^2$ for the first, second and third sample,
respectively. To convert the $\tau$ mass estimator $P_1$ to $M_\tau$, we use
the weighted mean and dispersion of $\Delta_1$, $\Delta_2$, and $\Delta_3$ to
obtain the estimator correction $\overline{\Delta}=(1.16\pm0.14)$~MeV/$c^2$.

The subtraction of this value from the edge position parameter $P_1$ in
data gives $M_\tau = 1776.61 \pm 0.13(\mathrm{stat.}) \pm 
0.14(\mathrm{MC}) $~MeV/$c^2$, where MC means the error due to limited
Monte Carlo statistics.

To study the systematic uncertainty due to the choice of the edge
parameterization, we use the following alternate functions:
\vspace*{-2mm}
\begin{equation}
F_1(x) = (P_3+P_4 x)\frac{x-P_1}{\sqrt{P_2+(x-P1)^2}}+P_5+P_6 x,
\label{eq:sqrt}\end{equation}
\vspace*{-5mm}
\begin{equation}
F_2(x) = (P_3+P_4 x)\frac{-1}{1+\mathrm{exp}((x-P_1)/P_2)}+
P_5+P_6 x
\label{eq:exp}\end{equation}
for the fit to the $M_\mathrm{min}$ distribution. Here $P_i$ are the
parameters of the fit.

The above procedure for $\tau$ mass extraction is repeated sucessively
for the data
and MC samples with each of these functions. The extracted values for
the $\tau$ mass obtained with the functions~(\ref{eq:sqrt}) and~(\ref{eq:exp})
are $(1776.85\pm0.13(\mathrm{stat.})\pm0.12(\mathrm{MC}))$~MeV/$c^2$ and
$(1776.52\pm0.12(\mathrm{stat.})\pm0.10(\mathrm{MC}))$~MeV/$c^2$, respectively.

We take for the measured value of the $\tau$ mass the one obtained using 
function (2):
\begin{equation}
M_\tau = M_1 = 1776.61 \pm 0.13 \mbox{(stat.) MeV/}c^2
\end{equation}
The square root of the variance of the obtained $\tau$ 
masses, 0.18 MeV/$c^2$ is taken as systematic
uncertainty due to the choice of the edge parameterization.
This value exceeds the error of 0.14 MeV/$c^2$, which is assigned
as a systematic uncertainty due to limited Monte Carlo statistics.
The use of a different fit range gives a much smaller shift in $\tau$ mass of
0.04~MeV/$c^2$, which we include in the systematic uncertainty.

In this analysis, we use the beam energy calibrated using the 
beam-energy constrained mass of fully reconstructed $B$ decays 
on a run-by-run basis. We estimate the uncertainty of the beam
energy to be less than 1.5 MeV, which includes the uncertainty of
$B$ mass, tracking system calibration and the effect of the 
$\Upsilon(4S)$ width.
Using the Monte Carlo samples, we find that this uncertainty 
propagates to a systematic uncertainty on the $\tau$ mass of 
0.26 MeV/$c^2$ under the assumption that the value 1.5 MeV is fully due to uncertainty
of the beam energy calibration.

As a cross-check of the result obtained from the fully reconstructed 
$B$ decays
we analyse the distribution of the variable
$\Delta_{ME} = (M(\mu^+\mu^-)-2E_\mathrm{beam})$ for
$e^+e^- \rightarrow \mu^+\mu^-$ data events. 
If a 
systematic shift exists in the
beam energy or tracking system calibration, we would expect some
shift of the maximum of this distribution from zero. A small shift of the
maximum of the $\Delta_{ME}$ distribution from zero is due to ISR and FSR. 
We fit the $\Delta_{ME}$ distribution to a sum of two 
Gaussians with the same central value 
multiplied by a cubic polynomial to
take into account the peak asymmetry
due to ISR. To check the consistency of this fitting procedure, we apply 
it to Monte Carlo $e^+e^- \rightarrow \mu^+\mu^-$ events with ISR and FSR
~\cite{KKMC} that pass through the full Belle simulation and reconstruction
procedures. The $\Delta_{ME}$ distributions
for data and Monte Carlo are shown in Fig. 2 together with results of the fit.
The reduced goodness-of-fit values $\chi^2/Ndf$ are 0.9 and 1.06 
for the data and Monte Carlo, 
respectively, where $Ndf=51$ is the number of
degrees of freedom.

\begin{figure}
\vspace*{-8mm}
  \includegraphics[width=0.4\textwidth] {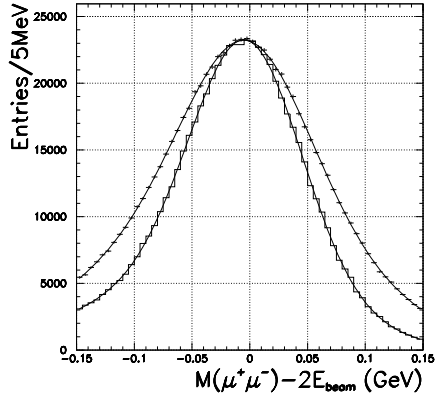}
\caption{The $(M(\mu^+\mu^-)-2E_\mathrm{beam})$ distributions 
for the data (points with errors) and Monte Carlo (histogram without errors).
The curves show the results of the fit to the data and Monte Carlo by the 
sum of
two Gaussians multiplied by a cubic polynomial.} 
\label{fig2}
\vspace*{-3mm}
\end{figure}

While the resolution of the $\Delta_{ME}$ variable is not
well described by Monte Carlo, the peak position coincides for data and
simulation. To estimate the systematics due to the difference in momentum
resolution between data and Monte Carlo we included additional
smearing of the track momenta in the Monte Carlo samples by a Gaussian
with $\sigma=1.02\cdot 10^{-3}\,p^2$ (the units of $\sigma$ and $p$ are
GeV/$c$). The consistency in $\Delta_{ME}$
between data and Monte Carlo becomes much better after this procedure.
The shift in the edge position parameter $P_1$ is
negligible (less than 0.02 MeV/$c^2$) and is included in total
systematics.   

The difference between data and Monte Carlo peak positions
obtained from the fit is $\delta \Delta_{ME} = 3\pm2$ MeV. This 
difference comes from the imperfect calibration of
both the beam energy and tracking system. 
We analyse two extreme cases when the shift $\delta \Delta_{ME}$ is due to
the imperfect calibration of either 1) the beam energy or 2) 
the tracking system.

For the first case, 
we have $\Delta E_\mathrm{beam} = \delta \Delta_{ME}/2 = 1.5$ MeV, which is
consistent with accuracy of the beam energy calibration obtained from the
reconstruction of the exclusive $B$ decays. To estimate the shift of the 
$\tau$ mass for the second case, we construct the $M_\mathrm{min}$ 
Monte Carlo distributions for an input $\tau$ mass equal to 1777.0 
MeV/$c^2$
for unmodified pion momenta and for momenta shifted by 
$\Delta p/p = \pm 3/10580 =  2.8 \times 10^{-4}$. 
We obtain a mass shift 
in the range 0.10--0.15 MeV/$c^2$, which is smaller than 
the shift observed when $\delta\Delta_{ME}$ includes the full
beam energy uncertainty (0.26 MeV/$c^2$). We take this 
conservative assumption and assume
a systematic uncertainty due to the combined imperfections of the 
beam energy and tracking system calibration of 0.26 MeV/$c^2$.        

To estimate the systematic uncertainty due to the model dependence
of the spectrum of the $3\pi$ system in the $\tau$ decay we vary
the mass and width of the $a_1(1260)$ meson in the 
range $\pm$ 300 MeV/$c^2$ from
the nominal PDG values. We find the shift in the edge position
due to this variation to be negligible (less than 0.02 MeV/$c^2$).
 
\begin{table}[ht]
\caption{Summary of systematic uncertainties}
\begin{tabular}{l|r}
{ Source of systematics} & { $\sigma,$~MeV/$c^2$} \\ \hline
Beam energy and tracking system & 0.26 \\
Edge parameterization & 0.18 \\
Limited MC statistics & 0.14 \\
Fit range & 0.04 \\
Momentum resolution &  0.02 \\
Model of $\tau \rightarrow 3\pi\nu_\tau$ & 0.02 \\
Background & 0.01 \\ \hline
Total & 0.35 \\
\end{tabular}
\end{table}

Systematic uncertainties from misidentified $\tau$
decay products and from non-$\tau^+\tau^-$ events are negligible 
(less than 0.01 MeV/$c^2$), 
since their $M_{\mathrm{min}}$ distributions show no
significant structure in the region of the $\tau$ mass.

The list of the analysed sources of systematics is given in Table I.
The final result is  
$M_\tau = (1776.61 \pm 0.13\mbox{(stat.)} \pm 0.35\mbox{(sys.))}$~MeV/$c^2$.
In the analysis we assume that the neutrino mass is equal to
zero. According to MC, a change in the neutrino mass from zero to
10 MeV/$c^2$ leads to a shift in the edge position of the pseudomass
distribution by $-0.1$ MeV/$c^2$.

The pseudomass method allows a separate measurement of the
masses of the positively and negatively charged $\tau$ leptons.  
A mass difference between positive and negative $\tau$ leptons would 
result in a difference in the energy between the $\tau$'s produced in the 
$e^+e^-$ collision.
This in principle makes the assumption $E_\tau = E_{\mathrm{beam}}$ invalid.
The $M_{\mathrm{min}}$ distributions for positive and negative $\tau$'s
decaying into $3\pi\nu_\tau$ are shown in Fig. 3 together with the results
of the fit. 

Good agreement is found between the distributions for $\tau^+$
and $\tau^-$. The mass difference obtained from independent fits
to these distributions is $M_{\tau^+}-M_{\tau^-} = (0.05\pm 0.23)$~MeV/$c^2$.

Most sources of systematic uncertainty on $\tau$ mass 
affect positive and 
negative $\tau$ leptons equally, so that their contributions
to the mass difference (and its uncertainty) cancel. 
One exception is different interactions of 
particles and antiparticles in the detector material. 
For example, the numbers of
positive and negative triplets for the selected 
pions are not equal to each other.
However, the description of this difference by the MC is 
reasonably accurate. In the
data the ratio of the number of negative to number of positive triplets is
1.034 while in MC this ratio is equal to 1.031. 
To estimate a systematic shift  in the mass difference 
between $\tau^+$ and $\tau^-$
we compare the
peak positions of 
$\Lambda_c \rightarrow pK^-\pi^+$ and 
$\overline{\Lambda}_c \rightarrow \bar{p}K^+\pi^-$,    
$D^+ \rightarrow \phi(1020)\pi^+$ and $D^- \rightarrow \phi(1020)\pi^-$,
$D_s^+ \rightarrow \phi(1020)\pi^+$ and 
$D_s^- \rightarrow \phi(1020)\pi^-$.
The average relative mass shift from the
decay modes listed above is approximately $0.8 \times 10^{-4}$.
This value is used as the systematic uncertainty in the relative
mass difference between $\tau^+$ and $\tau^-$ and corresponds to 
a systematic uncertainty in the mass difference  
of 0.14 MeV/$c^2$.

Adding the statistical and systematic errors in quadrature, we obtain
$M_{\tau^+}-M_{\tau^-} = (0.05\pm 0.27)$~MeV/$c^2$.
This result can be expressed as an upper limit on the relative mass 
difference~\cite{FC}
$|M_{\tau^+}-M_{\tau^-}|/M_{\tau} < 2.8 \times 10^{-4} 
~~\mbox{at} ~90\% ~\mbox{CL}$. 
Good agreement of the $M_{\mathrm{min}}$ distributions  
for positive and negative $\tau \rightarrow 3\pi\nu_\tau$ decays shows that 
CPT invariance is respected at the present
level of experimental accuracy.

\begin{figure}
\vspace*{-8mm}
  \includegraphics[width=0.4\textwidth] {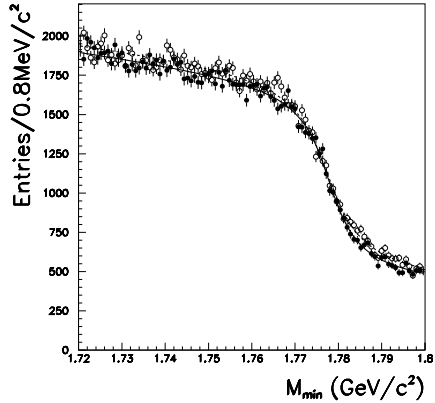}
\caption{The distribution of the pseudomass $M_{\mathrm{min}}$ for
the decay $\tau \rightarrow 3\pi^\pm \nu_\tau$, shown separately for
positively and negatively charged $\tau$ decays.
The solid points with error bars correspond to
$\tau^+$ decays, while
the open points with error bars are $\tau^-$ decays.
The solid curve is the result of the fit to the pseudomass distribution of $\tau^+$
with function (2) while the dashed one is for the $\tau^-$. }
\label{fig3}
\vspace*{-3mm}
\end{figure}

\medskip
To summarize, we have measured the mass of the $\tau$ lepton from the pseudomass
distribution of $\tau$ decays into three charged pions and a neutrino.
The result is
\begin{displaymath}
M_\tau = (1776.61 \pm 0.13\mbox{(stat.)} \pm 0.35\mbox{(sys.))} ~\mbox{MeV/}c^2,
\end{displaymath}
in good agreement with the current world average and of comparable
accuracy. Independent measurements of the positive and 
negative $\tau$ mass are obtained to test CPT symmetry. 
The measured values are 
consistent and an upper limit on the relative mass difference is
\begin{displaymath}
|M_{\tau^+}-M_{\tau^-}|/M_{\tau} < 
 2.8 \times 10^{-4} ~~\mbox{at} ~90\% ~\mbox{CL}, 
\end{displaymath}
one order of magnitude better than the previous limit from OPAL.

We thank the KEKB group for excellent operation of the
accelerator, the KEK cryogenics group for efficient solenoid
operations, and the KEK computer group and
the NII for valuable computing and Super-SINET network
support.  We acknowledge support from MEXT and JSPS (Japan);
ARC and DEST (Australia); NSFC and KIP of CAS (contract No.~10575109 
and IHEP-U-503, China); DST (India); the BK21 program of MOEHRD, and the
CHEP SRC and BR (grant No. R01-2005-000-10089-0) programs of
KOSEF (Korea); KBN (contract No.~2P03B 01324, Poland); MIST
(Russia); ARRS (Slovenia);  SNSF (Switzerland); NSC and MOE

\end{document}